# Most nearby young star clusters formed in three massive complexes

Cameren Swiggum[1], João Alves[1], Robert Benjamin[2], Sebastian Ratzenböck[1,3], Núria Miret-Roig[1], Josefa Großschedl[1,4], Stefan Meingast[1], Alyssa Goodman[5], Ralf Konietzka[5], Catherine Zucker[5], Emily L. Hunt[6], Sabine Reffert[6]

[1] University of Vienna, Department of Astrophysics, Türkenschanzstraße 17, 1180 Wien, Austria

[2] Department of Physics, University of Wisconsin-Whitewater, 800 West Main St, Whitewater, WI 53190, USA

[3] University of Vienna, Research Network Data Science at Uni Vienna, Kolingasse 14-16, 1090 Vienna, Austria

[4] Physikalisches Institut, Universität zu Köln, Zülpicher Str. 77, D-50937 Köln, Germany

[5] Center for Astrophysics | Harvard & Smithsonian, 60 Garden St. Cambridge, MA 02138, USA

[6] Landessternwarte, Zentrum für Astronomie der Universität Heidelberg, Königstuhl 12, 69117 Heidelberg, Germany

**Efforts to unveil the structure of the local interstellar medium and its recent star formation history have spanned the past seventy years[1–6]. Recent studies utilizing precise data from space astrometry missions have revealed nearby, newly formed star clusters with connected origins[7–12]. Nonetheless, mapping young clusters across the entire sky back to their natal regions has been hindered by a lack of clusters with precise radial velocity data. Here we show that 155 out of 272 (57 percent) high-quality young clusters [14,15] within one kiloparsec of the Sun arise from three distinct spatial volumes. This conclusion is based upon the analysis of data from the third Gaia release[13] and other large-scale spectroscopic surveys. Currently dispersed throughout the Solar Neighborhood, their past positions over 30 Myr ago reveal that these families of clusters each formed in one of three compact, massive star-forming complexes. One of these families includes all of the young clusters near the Sun – the Taurus and Sco-Cen star-forming complexes[16,17]. We estimate that over 200 supernovae were produced from these families and argue that these clustered supernovae produced both the Local Bubble[18] and the largest nearby supershell GSH 238+00+09[19], both of which are clearly visible in modern three-dimensional dust maps[20–22].**

We have used a recently published star cluster catalog[14] to create a sample of 254 high-quality, young (< 70 Myr) clusters within approximately one kiloparsec of the Sun, supplemented with an additional set of 18 Young Local Associations[15] (YLAs), which are low-mass, co-moving associations of stars located within 200 parsecs of the Sun. Here we refer to the members of both samples as 'clusters'. Using the current positions and three-dimensional space velocity of each cluster, we integrate the clusters' orbits 60 Myr backwards in time assuming an axisymmetric Galactic potential[23]. Going back 30 to 50 million years, we find that nearly 60% of these clusters' trajectories converge at three locations, indicating that a large fraction of clusters in the solar neighborhood share common origins. We apply the HDBSCAN[24] algorithm to each timestep of the clusters' past positions based on their computed orbits to create a membership list for each family. We designate the three cluster families as the Collinder 135 (Cr135), Messier 6 (M6), and Alpha Persei ($\alpha$Per) families, named after the most notable old cluster in each grouping. These families contain 39, 34, and 82 clusters respectively, accounting for 57% of the 272 clusters in our sample and 59% of the 48,514 stars in our sample. We show the all-sky distribution of each family's stars in Figure 3 (Supplementary Fig. 3).

Figure 1 (Supplementary Fig. 1) shows the positions of all 272 clusters as a function of time from 60 Myr ago to the present; the clusters belonging to the Cr135, M6, and $\alpha$Per families are highlighted with different colored dots, with smaller dots for lookback times that are greater than the estimated cluster age. We see these families were previously much more compact. At each time step, we determine a family's 3D size by calculating the median distance between individual clusters and their respective geometric center. These values are reported in Table 1 and shown in Extended Data Figure 1. The most compact size of the Cr135, M6, and $\alpha$Per families were $49^{+6.2}_{-9.7}$ pc ($30.9^{+2.2}_{-1.2}$ Myr ago), $76.6^{+4.0}_{-15.9}$ pc ($37.4^{+9.3}_{-4.6}$ Myr ago), and $88.3^{+10.4}_{-15.4}$ pc ($20.0^{+2.5}_{-2.9}$ Myr ago), respectively. The Cr135 family has since expanded by a factor of 4.6, while the M6 and $\alpha$Per families have expanded by a factor of 2.3 and 2.1, respectively. A significant fraction of clusters in each family were formed around or following the time these families were at their minimum size. Nonetheless, the star formation histories for these families are prolonged (Extended Data Figure 2), with the M6 and $\alpha$Per groups forming stars for a duration of roughly 60 million years.

The three panels of Figure 2 (Supplementary Fig. 2) show the current XY position of each family overlaid on a kiloparsec-scale 3D dust map[21]. The clusters of the three families appear to be associated with regions of low dust density. The interactive version of Figure 2 displays the 3D cluster positions on a dust map and a model of the Local Bubble (LB)[25]. The clusters of the $\alpha$Per are predominantly situated within and around the LB – a supernova driven cavity with a shell mass of $1.4 \times 10^6 M_\odot$ and a radius of 165 pc, which is likely responsible for the formation of the Taurus and Sco-Cen star-forming regions on its surface[18]. The Sun is located near the center of the LB, which is at (x, y, z) = (39, 7, –18) pc[18]. Most of the Cr135 family and 15 M6 family clusters lie within a kiloparsec-long, 600 pc-wide Galactic supershell, GSH 238+00+09 (hereafter GSH 238), identifiable in HI, dust, and X-rays with a previously estimated HI mass of $2.7 \times 10^6 M_\odot$[19]. The left boundaries of the Local Bubble (LB) and GSH 238 are part of the Radcliffe Wave (RW), a 2.7 kpc long, oscillating chain of molecular clouds shown in Figure 2[26,27]. The remaining M6 clusters span a dust gap between Sco-Cen and Vela Molecular Ridge into low-density areas of the first and fourth Galactic quadrants. We use stellar isochrones and assumptions for the initial mass function (IMF)[28] and star-formation efficiency to estimate the total stellar mass of the clusters comprising each family, their initial gas masses, and the number of possible supernovae (SNe) progenitors. The results are given in Table 1 and the most notable members of each family are listed in the Methods. A total of 62, 125, and 61 SNe, mostly occurring over the past 30 Myr, are expected to be associated with the Cr135, M6, and $\alpha$Per families, respectively. The initial gas masses of each family's progenitor star-forming complex are on the order of $10^6 \, M_\odot$.

The Collinder 135 family includes several prominent clusters, such as Collinder 135, NGC 2547, and IC 2395, each of which are displayed in Figure 3. The youngest members tend to be situated farthest from the family's center, including ASCC 18 and ASCC 20 within the Orion complex[6,29] and IC 2395, part of the Vela OB1c association[30]. The six clusters extending the furthest into the third Galactic quadrant compose the Collinder 121 (CMa OB2) OB association[31]. The connection between the GSH 238 super-shell and the Cr135 family is supported by an examination of the size and kinematics of the shell. Using the measured expansion velocity[19] and size of GSH 238, combined with an analytic model of shell expansion in a uniform medium driven by supernovae[32], we calculate the shell's dynamical age to be 29.4 ± 4.3 Myr, requiring 96 ± 40 supernovae. This agrees with the average age of the clusters of the Cr135 family (approximately $21.7^{+10.5}_{-8.6}$ Myr), providing additional evidence that the clusters in the family are responsible for the formation of the shell. A relationship between the family's Cr135, Cr140, and NGC 2451b clusters was recognized before space astrometry missions[33,34]. Recently, *Gaia* data helped reveal connected structure within the Cr135 family, although the lack of *Gaia* DR3 radial velocity data previously hindered tracing their compact origins[8,9,11,12].

The Messier 6 (M6) family spans the third and fourth Galactic quadrants and includes prominent clusters such as Messier 6 (M6 or NGC 6405), Trumpler 10, IC 2391 ('Omicron Velorum cluster'), NGC 2451A, and NGC 3228, all depicted in Figure 3. It is the most massive of the three families and likely produced the most supernovae (124). The family's clusters mainly occupy dust cavities shown in Figure 2, spreading across the third and fourth Galactic quadrants and within GSH 238, rather than a single, defined dust shell. Ten clusters of the M6 family, which we estimate have produced around 28 SNe, began to overlap with the Cr135 family approximately 15 million years ago. This brings the number of SNe potentially contributing to the expansion of GSH 238 to 90, closer to the 96 supernovae needed to drive the observed expansion of the supershell. Prior works suggested that Trumpler 10 and Haffner 13 formed from the same star-forming complex with notable clusters of the Cr135 family[8,11,12]. Figure 1 shows that the M6 family clusters, including Trumpler 10 and Haffner 13, were spatially separated from the Cr135 family during their formation and have only begun to occupy the same volume over the past 15 Myr.

The Alpha Persei ($\alpha$Per) family includes many clusters historically linked to nearby interstellar structure seen in studies of gas and dust[1–6]. It includes notably the $\alpha$Per cluster (Melotte 20), IC 2602 ('Southern Pleiades'), and IC 4665, which are highlighted in Figure 3. Most of the clusters in the $\alpha$Per family are located within or along the edge of the Local Bubble. A recent investigation found that the clusters of the Sco-Cen and Taurus complexes, members of the $\alpha$Per family, formed along the bubble's edge and are moving away from its center[18]. In this model, the Local Bubble originated 15–30 Myr ago from SNe occurring in UCL and LCC; IMFs of these clusters indicate a supernova production of 14-20 SNe[35–37]. This agrees with a momentum analysis of the bubble's expansion, which requires $15^{+11}_{-7}$ SNe[18]. Our work confirms that clusters within and near the UCL and LCC regions have generated approximately $11^{+4}_{-4}$ SNe. These findings connect the older $\alpha$Per cluster with the young clusters on the LB, affirming a previous hypothesis that feedback from the $\alpha$Per cluster initiated a sequence of star formation, eventually resulting in the LB's associated star-forming regions[2,6,38]. Recent work [39] which posited that the Sco-Cen complex extends through clusters IC 2391, NGC 2541A, and Cr135 is not supported by our findings, which show these clusters have distinct origins from Sco-Cen.

Our study uncovers the origins of many young local star clusters and clearly identifies them as major influences in shaping the local interstellar medium. This result provides a framework for studying the effects of star-formation feedback on the interstellar medium of galaxies and opens the possibility of searching for other cluster families, and their effects, across the Milky Way.

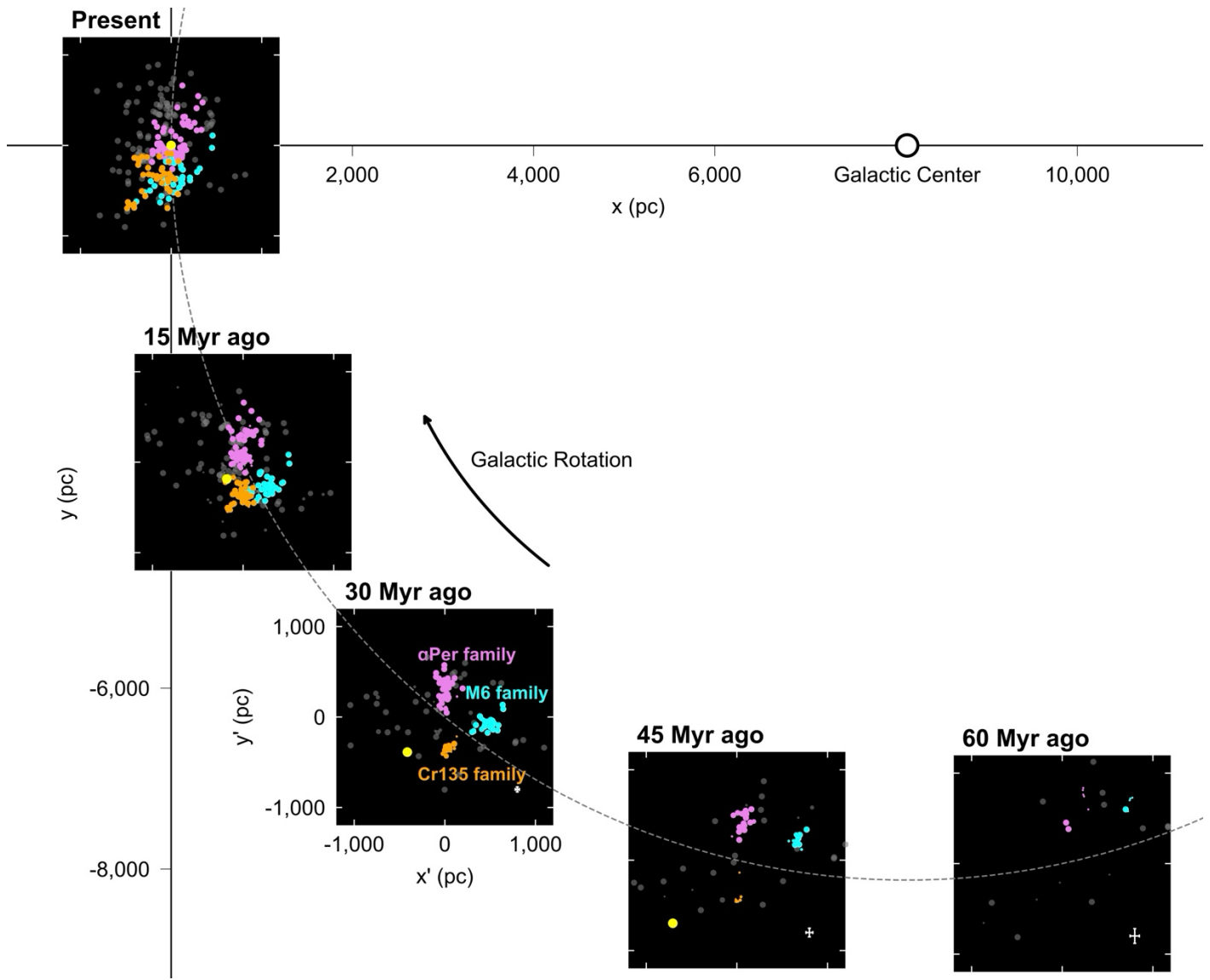

**Fig 1 ([INTERACTIVE](#)): A Galactic bird's-eye (*XY*) view of the clusters' orbits over time,** showing over one quarter of a circle (gray, dashed line) of radius of $R_{Sun}$=8,122 pc extending from the Galactic Center to the Sun. The location of the Galactic Center and the direction of Galactic rotation are indicated. The five square panels show the clusters' positions as dots at five different times: the present (upper left), 15, 30, 45, and 60 million years ago (from left to right). The center of each panel shows the location of the local standard of rest (LSR) at each time. The three-dimensional [interactive](#) version of this figure has a time-slider showing the cluster positions at intermediate time steps with each frame centered on the LSR. The clusters of each family are color-coded and labeled; the remaining clusters are shown as gray dots. Dots become smaller and eventually disappear at times older than the age of a cluster. White crosses in the lower right-hand corner of the final three panels show the cluster sample's median *x* and *y* standard deviations when tracing back the orbits: approximately 25 pc at *t* = -30 Myr, 45 pc at *t* = -45 Myr, and 50 – 80 pc at *t* = –60 Myr, from 100 orbit realizations. Note that over a span of 45 million years, each of the three cluster families transitions from a more dispersed to a tighter configuration.

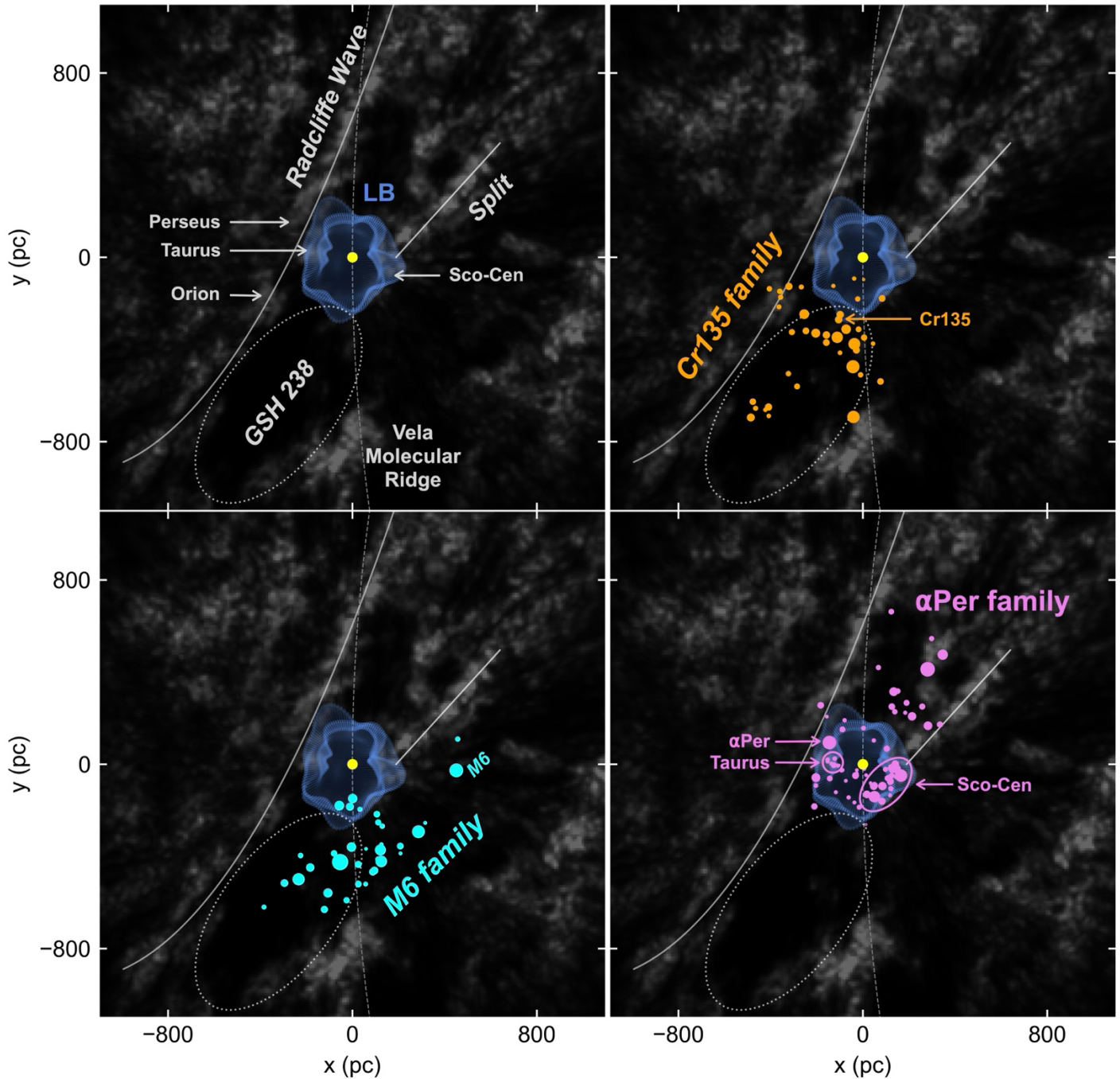

**Figure 2 (INTERACTIVE): A bird's-eye (*XY*) view of the present-day positions of galactic clusters overlaid on a kiloparsec-scale, 3D dust map (grayscale).** Each of the four panels shows the same region centered on the Sun's position (yellow dot). The solid white lines show the extent of the two major dust features found in the local Milky Way, the Radcliffe Wave and the Split, as well as arrows indicating the locations of nearby star-forming complexes. The panels also show a fit to the boundary of the Local Bubble (LB; outlined in blue) and the kiloparsec long dust shell, GSH 238+00+09 (GSH 238; dotted white ellipse). The upper-right, lower-left, and lower-right panels display the isolated positions of three principal families of clusters, with symbol sizes proportional to their stellar masses. Collinder 135 (Cr135) clusters are predominantly located within the GSH 238 shell, with some also extending into the LB. M6 family clusters are partially located within the GSH 238 shell and the Local Bubble but also extend towards positive *XY* into unnamed regions of lower dust density. Alpha Persei (αPer) family clusters are mostly associated with the Local Bubble, with extensions towards positive *XY*, particularly along and near the Split. The interactive version of the figure shows a comparison of the dust distribution and cluster positions in three dimensions.

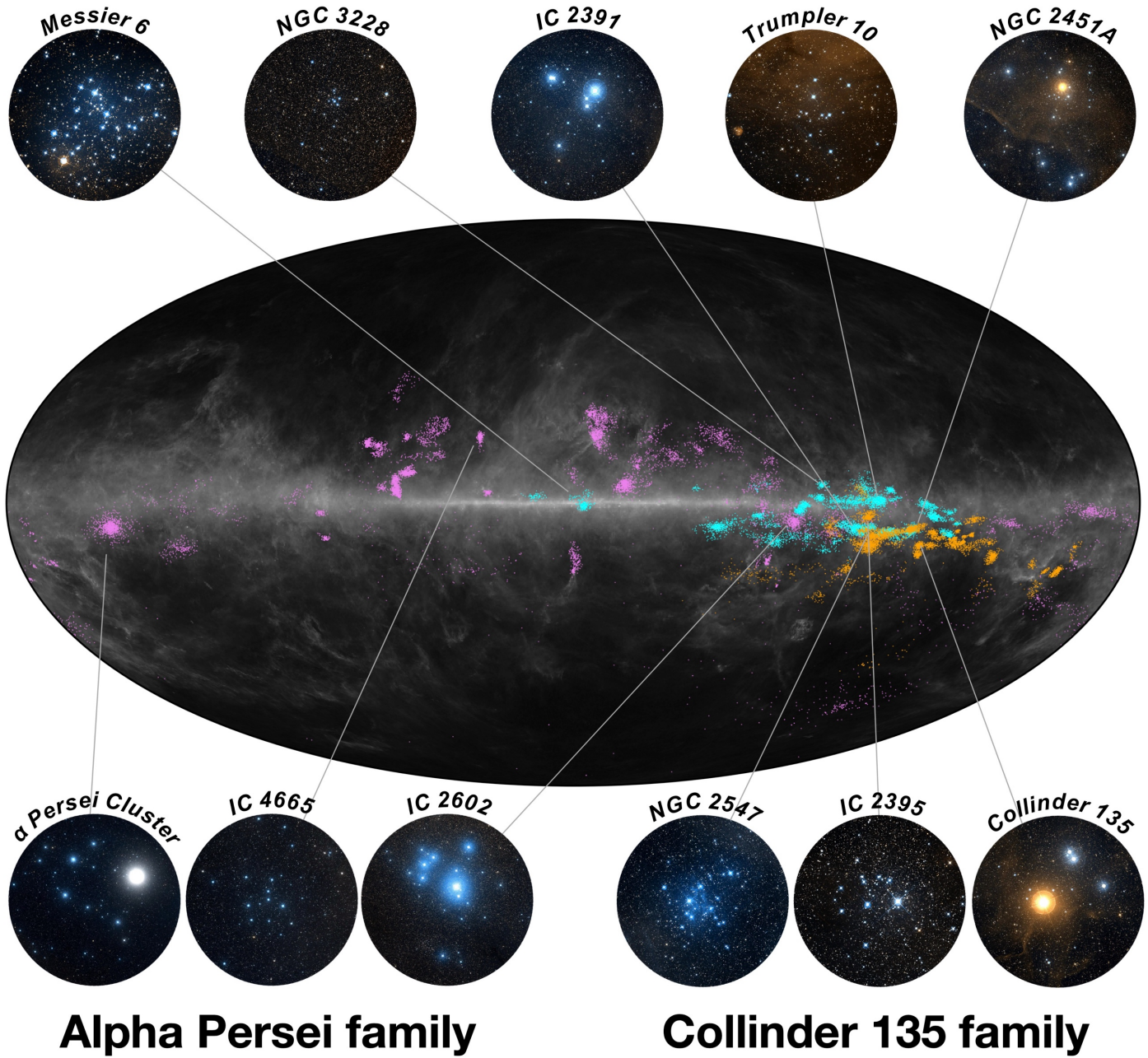

**Figure 3 (INTERACTIVE): All-sky positions of the clusters' stars and high-resolution optical images of the most notable clusters.** The background gray-scale image is the *Planck* 853 GHz map tracing dust emission. The colored points show the Galactic sky positions (*l,b*) of *individual stars* belonging to the three cluster families. The circles surrounding the all-sky plot display optical images of the notable clusters taken from the Digitized Sky Survey II and the thin, gray lines connect these images to their locations on the sky. Since the clusters of the αPer family are on average closer to the Sun than the other two families, their positions are more spread across the sky.

| Family | N | l (deg) | b (deg) | Age (Myr) | $N_{SNe}$ | Stellar Mass ($\times 10^3 M_\odot$) | Gas Mass ($\times 10^6 M_\odot$) | $D_{present}$ (pc) | $D_{compact}$ (pc) | $t_{compact}$ (Myr) | $v_{exp}$ (km s$^{-1}$) |
|---|---|---|---|---|---|---|---|---|---|---|---|
| Cr135 | 39 | 248.9 | -11.1 | $21.7^{+10.5}_{-8.6}$ | $63^{+8}_{-8}$ | $10.4^{+0.3}_{-0.3}$ | 0.7 – 2.1 | $226.6^{+23.3}_{-37.7}$ | $49.0^{+6.2}_{-9.7}$ | $-30.9^{+1.2}_{-2.2}$ | $4.2^{+8.3}_{-2.0}$ |
| M6 | 34 | 356.6 | -0.8 | $32.4^{+13.8}_{-10.5}$ | $124^{+11}_{-9}$ | $14.1^{+0.5}_{-0.5}$ | 0.9 – 2.8 | $178.2^{+24.8}_{-29.3}$ | $76.6^{+4.0}_{-15.9}$ | $-37.4^{+4.6}_{-9.3}$ | $3.1^{+5.4}_{-2.0}$ |
| αPer | 82 | 147.4 | -6.4 | $19.8^{+24.5}_{-13.5}$ | $61^{+7}_{-7}$ | $10.6^{+0.3}_{-0.3}$ | 0.7 – 2.1 | $181.3^{+18.1}_{-19.0}$ | $88.3^{+10.4}_{-15.4}$ | $-20.0^{+2.9}_{-2.5}$ | $3.3^{+2.6}_{-2.6}$ |

**Table 1**: **Computed properties of each family** including number of clusters (column 2), galactic longitude and latitude of the namesake cluster (column 3 and 4), age of clusters (column 5), inferred number of supernovae explosions (column 6), inferred stellar mass (column 7), range in estimated initial gas mass (column 8), present-day size (column 9), most compact size (column 10), the time of most compact size (column 11), and the expansion velocity (column 12). Quantities are reported as median estimates, with lower and upper boundaries representing the differences between the median and the 16th percentile, and the 84th percentile and the median, respectively. We note that the lower and upper bounds in the ages (column 5) and expansion velocities (column 12) represent the spreads in these quantities across a family and are not meant to indicate the uncertainty.

# Methods

## Construction of cluster sample

We use a *Gaia* DR3[13] based catalog[14] of 7,167 star clusters constructed with the Hierarchical Density-Based Spatial Clustering of Applications with Noise algorithm (HDBSCAN)[24], an unsupervised clustering routine that relies minimally on user-input parameters. The catalog contains gravitationally bound globular clusters, open clusters, and unbound moving groups. Our analysis uses several cluster parameters from this catalog, including sky position, distances, proper motions, heliocentric Galactic cartesian coordinates $(x,y,z)$, and the derived ages. We initially restrict our sample to clusters that meet certain quality criteria, *cst* > 5 and *class_50* > 0.5, reducing the sample to 4,105 clusters. We further restrict our analysis to 1,250 remaining clusters near the Sun: $-1,000 < x,y < 1,000$ pc and $-300 < z < 300$ pc. This sample has median uncertainties in $(x,y,z)$ of $\pm$ (1.2, 1.4, 0.2) pc. Since we are interested in studying recent generations of star formation, we further limit the sample to the 447 clusters with estimated ages less than 70 Myr. This age threshold, as implemented previously[10], is chosen because of the growing uncertainties in a cluster's trajectory beyond 70 Myr. We crossmatch the stars of these 447 clusters to the APOGEE-2 DR17[40] and GALAH DR3[41] surveys by matching the reported *Gaia* DR3 source ID to obtain radial velocities (RVs) in addition to the provided *Gaia* DR3 RVs. We remove RVs with reported errors greater than 5 km s$^{-1}$. If a star has an RV recorded in multiple of the three surveys, we adopt the RV with the lowest percentage uncertainty. For each star, we compute the heliocentric cartesian velocities $U, V, W$ and remove stars in each cluster with $U, V,$ or $W$ values more than $5\sigma$ away from the cluster median; this helps eliminate stars with erroneous RVs, such as spectroscopic binaries. We compute the median value and uncertainties in $U, V, W$ for each cluster. Clusters with fewer than five RV measurements or where the uncertainties in a cluster's bulk $U, V,$ or $W$ velocities exceed 5 km s$^{-1}$ are removed yielding a final sample of 254 clusters consisting of 47,670 stars.

We supplement this with a catalog of 27 Young Local Associations (YLAs)[15], which are low-mass, coeval moving groups of stars within 200 pc from the Sun. We limit our sample of this catalog to all YLAs younger than 70 Myr, except IC 2391, IC 2602, and Platais 8, which were already included in the sample above. This leaves 18 YLAs consisting of 639 stars. For each of these YLAs, we use their average spatial coordinates and velocities which were previously determined through a multivariate Gaussian-fitting approach applied to their stellar distributions based on *Gaia* DR1 astrometry. We also adopt the corresponding co-variances from this procedure to represent the errors associated with these measurements. Given the proximity of YLAs, no other cuts in the sample were made.

Combining the sample of 254 clusters and 18 YLAs yields a final sample of 272 young star clusters (48,309 stars) for our analysis. These clusters have a median age of $21.5^{+23.0}_{-14.0}$ Myr, median uncertainties in (*x, y, z*) of ± (0.3, 0.5, 0.1) pc and median uncertainties in (*U, V, W*) of ± (0.7, 0.9, 0.2) km s$^{-1}$.

The results described in this work were originally found using a compilation of seven previously published cluster catalogs constructed with *Gaia* data, supplemented with radial velocities as described above. The release of the catalog used here superseded this preliminary effort since the clusters have a uniform construction using HDBSCAN and uniformly determined ages. Our preliminary work indicates that the results reported here are independent of the methods used to compile the cluster catalog.

## Orbit Integration

We use the galactic dynamics package *galpy*[23] to calculate the past trajectory of each cluster using the *MWPotential2014* potential for the combination of the Galactic halo, bulge, and disk. We scale the model to a circular velocity of 236 km s$^{-1}$, with a solar Galactocentric radius of $R_\odot$ = 8.122 kpc[42] and vertical position of $z_\odot$ = 20.8 pc[43]. We use the initial positions and velocities of the 272 clusters described above as input into the *galpy* 'Orbit' module and use the *scipy odeint* procedure to integrate their orbits 60 Myr into the past using a timestep of 0.1 Myr. In order to calculate orbits, *galpy* converts the heliocentric Cartesian velocities into Galactocentric by subtracting an adopted peculiar motion of the Sun: ($U_\odot$, $V_\odot$, $W_\odot$) = (11.1, 12.24, 7.25) km s$^{-1}$ [44]. The 60 Myr traceback limit is adopted to match our chosen age limit for the cluster sample and avoid the growing uncertainties in trajectories with increasing look-back time[10]. For timesteps preceding the age of a given cluster, we calculate the "pre-birth" trajectory of the cluster as an approximate indicator of the parent gas cloud motion. Since it is likely that the parent gas cloud was subject to additional forces owing to pressure gradients, these pre-birth trajectories should be regarded with caution. Alterations in the gravitational potential from a spiral arm transit, encounters with GMCs[45], and interactions with other star clusters might have affected the clusters' trajectories and are not considered in our analysis. We consider these effects to be negligible given the relatively short time-period of our tracebacks[10].

We construct normal distributions for all six coordinates characterizing each cluster's present-day position and velocity and generate one hundred different realizations for each cluster's orbit. At *t* = –60 Myr, the cluster sample's median positional standard deviations are (*x,y,z*) = (51.1, 81.1, 3.6) pc. To visualize the cluster positions as a function of time in the interactive version of Figure 1, we define the coordinates (*x',y',z'*) to be the position of each cluster with respect to the motion of the Local Standard of Rest using the circular velocity above.

We have also integrated the orbits of the clusters using two alternate Galactic potential models provided in the *galpy* package. In the first model[46], the Sun has a circular velocity of 233 km s$^{-1}$, with $R_\odot$= 8.21 kpc. In the second model[47], the Sun has a circular velocity of 242 km s$^{-1}$, with $R_\odot$= 8.4 kpc. In addition, we investigated the effect of different adopted values for the Sun's peculiar motion with respect to the Local Standard of Rest, using ($U_\odot$, $V_\odot$, $W_\odot$) = (10.0, 5.2, 7.2) km s$^{-1}$ [48] or (10.1, 15.4, 7.8) km s$^{-1}$ [49].

The derived properties of the identified cluster families and their statistical and systematic uncertainties are described after we discuss how the families are constructed.

## Uncovering cluster families with HDBSCAN

A visual examination of the trajectories for the full cluster sample (Figure 1; Supplementary Fig. 1) makes it apparent that many clusters originate from three prominent overdensities. The collection of clusters making up each overdensity are referred to as "families". We apply HDBSCAN to robustly identify the members of each family. HDBSCAN is run on the clusters' (*x,y,z*) positions at each timestep of our orbital integration, determining whether a given cluster is identified as a family member or a noise category at each timestep of its orbit.

For each timestep, we use HDBSCAN with *cluster_selection_method* set to 'eom' and with *min_cluster_size* set to values from 5 to 30, increasing in steps of 1. This range of *min_cluster_size* adopted is based on our visual inspection. For each cluster and for every value of *min_cluster_size*, we construct a set of labels with the assigned family membership or noise category. Determining a final set of labels (or consensus function) for each cluster requires solving an optimization problem that pairs up data points that appear for the same cluster for the different cases. We apply a consensus function specifically designed for density-based clustering with noise developed for previous investigations[17,50].

The family (or noise) a given cluster is assigned to by HDBSCAN may change for each time step depending on the position of a cluster along its orbit. To decide on a final label for each cluster, we take advantage of the estimated cluster ages. For each cluster, we weigh the contribution of labels differently for different time steps. The weight function is a Gaussian that is centered on a cluster's estimated age and has a standard deviation of 20 Myr (for all clusters). The label with the largest total weight, integrated over all timesteps, is the adopted label for the cluster; the total weight for the adopted cluster label is recorded. For clusters which are labeled as noise less than 90% of the time, we adopt the label with the second largest weight. This modification captures clusters that originate in the visually identified overdense regions but migrate away so rapidly that they are labeled as noise most of the time. This adjustment runs the risk of including clusters that were just migrating through the overdense region; for these clusters, we also visually examine their orbits with respect to clusters with more secure family memberships. This approach is repeated for all 100 realizations of the cluster trajectories. We then average the weights assigned to each label for all realizations to produce a "weight score" for each cluster. Each of the families that are identified using this approach are assigned a "mean weight score" defined as the mean of the aggregated weight of the individual family members for all 100 realizations.

Our approach successfully recovers the three families of clusters that were visually identified as overdensities, along with a fourth, smaller family. We name them the Collinder 135 (Cr135) family, the Alpha Persei ($\alpha$Per) family, the Messier 6 (M6) family. These families and a smaller, unnamed family, initially consist of 43, 96, 43, and 8 clusters, respectively, and have mean weight scores of 0.45, 0.51, 0.55, and 0.29. Each family is named after the most notable old cluster in each grouping. We see visual evidence for additional smaller families, but these were not recovered by our algorithm due to the range of HDBSCAN *min_cluster_size* that we adopted for this work.

For each family, we visually examine the orbits of its constituent clusters. As expected, most of their member clusters move together coherently, but the large families seem to contain some interlopers which do not move consistently with their sibling clusters. We examine these interlopers and compare their individual weight scores to the mean weight score of their family. The Cr135 family has four interlopers each with weight scores below the mean weight score of the family. We remove these clusters from the Cr135 family "by hand", leaving it with 39 members. The M6 family includes nine visual interlopers. All but one of these clusters have weight scores below the family mean. The one cluster with a higher weight score (HSC 1677) moves with the bulk of the M6 family but formed approximately 200 pc away from the other clusters that formed at this time (60 Myr ago). We remove all nine of these interloper clusters from the M6 family, leaving it with 34 clusters. The $\alpha$Per family contains the largest number of interlopers: 14 clusters. Six of these interloper clusters (CWNU 1183, HSC 633, RSG 5, Theia 96, CWNU 519, HSC 782) appear to move collectively in the direction opposite to the other $\alpha$Per family clusters. Three of these six clusters have weight scores greater than the family value since they spend a considerable fraction of their orbits passing through the overdensity of the reliable $\alpha$Per family members. We remove these six clusters, along with the other eight interloper clusters with weight scores below the family mean, leaving a total of 82 clusters in the $\alpha$Per family. The source data table (Supplementary Information) lists all 272 clusters, their family membership, and whether they were visually identified as interlopers and removed from their initially assigned family.

In this manuscript we focus our analysis on the newly identified Cr135, M6, and $\alpha$Per families. The smaller family (eight clusters) may have some relation to the M6 family; this will be addressed in future work. All figures throughout this work color-code the Cr135 family as orange, the M6 family as cyan, and the $\alpha$Per family as violet.

The distributions of ages for the clusters comprising each family are given in Extended Data Figure 2. We list notable clusters of each family in the following three paragraphs. The entire cluster membership of the families can be found in the source data table (Supplementary Information).

The Collinder 135 family most notably includes the Octans association (OCT; 35 Myr), Collinder 135 (Cr135; 30 Myr), Alessi 36/UBC 7 (29 Myr), NGC 2451B (28 Myr), NGC 2547 (22 Myr), Collinder 140 (Cr140; 17 Myr), Collinder 132 (Cr132; 14 Myr), ASCC 20 (13 Myr), ASCC 18 (9 Myr), and IC 2395 (7 Myr). Other clusters of the family with more than 200 stars (approximately above the 75th percentile of the parent cluster sample) include FoF 2383 (38 Myr), CWNU 1024 (32 Myr), OC 0450 (22 Myr), Alessi 34 (21 Myr), and HSC 1865 (8 Myr).

The Messier 6 (M6) family's prominent members include NGC 6405/M6 (52 Myr), Trumpler 10 (37 Myr), Platais 9 (37 Myr), NGC 3228 (30 Myr), IC 2391 (27 Myr), NGC 2451A (26 Myr), and Haffner 13 (15 Myr). Other clusters (with more than 200 stars) include ASCC 58 (57 Myr), Alessi 5 (56 Myr), BH 99 (45 Myr), CWNU 1044 (34 Myr), CWNU 45 (33 Myr), LISC-III 3668 (29 Myr), Theia 58 (24 Myr), and BH 164 (19 Myr).

The Alpha Persei ($\alpha$Per) family consists notably of the $\alpha$Per cluster (Melotte 20; 56 Myr), Platais 6 (48 Myr), the Carina association (CAR; 45 Myr), the Tucana-Horologium association (THA; 45 Myr), the Columba association (COL; 42 Myr), IC 4665 (33 Myr), IC 2602 (26 Myr), Platais 8 (31 Myr), the $\chi^1$ association (XFOR; 30 Myr), IC 2602 (26 Myr), the β Pictoris association (BPMG; 24 Myr), and the 32 Orionis association (THOR; 22 Myr). The family includes the entire Sco-Cen complex, which is classically divided into three sub-populations[51,52], namely Upper Centaurus Lupus (UCL; 16 Myr), Lower Centaurus Crux (LCC; 15 Myr), and Upper Scorpius (USCO; 10 Myr). Sco-Cen is also host to younger clusters associated with regions of active star-formation, including Chameleon I (6 Myr), Corona-Australis (CRA; 4 Myr), η Chamaeleontis (ETAC or Mamajek 1; 4 Myr), ε Chamaeleontis (EPSC; 4 Myr), and ρ Ophiuchi (ROPH; 1 Myr). The $\alpha$Per family also includes five young clusters in the Taurus star-forming region. Other clusters of the family (with more than 200 stars) include HSC 1553 (38 Myr), HSC 381 (31 Myr), Stephenson 1 (Delta Lyrae cluster; 27 Myr), UBC 26 (25 Myr), HSC 2636 (10 Myr), HSC 2907 (10 Myr), V1062-Sco (UPK 640; 10 Myr), HSC 2468 (8 Myr), HSC 2986 (7 Myr), Theia 38 (6 Myr), and OCSN 96 (5 Myr).

**Measuring family properties: sizes, ages, expansion velocities, and radial drift**
For each of the cluster families, we calculate the following parameters: $D_{compact}$, the minimum size (pc) of the family, $t_{compact}$, the time (Myr) at which each family had its minimum size, $v_{exp}$, the expansion speed (km s$^{-1}$) of the clusters with respect to the family cluster-weighted barycenter and $\Delta R$, the drift (kpc) of the family in Galactocentric radius.

We define the family "center" at each timestep as the unweighted, median $(x',y',z')$ position of its constituent clusters. The "size" of each family, $D$ (pc), is the median distance between clusters of each family and the family center; this is plotted as a function of time for each family in Extended Data Figure 1. $D_{\text{compact}}$ is the minimum size for each family and $t_{\text{compact}}$ is the time at which it occurs. To obtain the uncertainties in these values, we generate 1000 bootstrapped samples by randomly drawing clusters from each family and re-computing $D_{\text{compact}}$ and $t_{\text{compact}}$. We calculate the median values for these quantities, compute their uncertainties as the 16th and 84th percentiles, and report them for each family in Table 1. We note that each family, especially the $\alpha$Per family, includes clusters that formed before their times of most compactness, as seen in Extended Data Figure 1. Extended Data Figure 2 shows that each family has extended star formation histories; clusters emerging in the initial epochs of a family's star formation history might have unique centers of convergence within the entire family volume that might not be immediately recognizable when examining the size of a family in its entirety. Future work is needed to clarify the full star formation histories of each of the three families, identify their possible sub-groups, and assess whether there are multiple points of convergence among sub-groups of clusters in a family.

To calculate the expansion speed of each family, we begin by calculating the median velocity of each family. This median velocity is then subtracted from the velocity of each cluster within the family, resulting in a velocity for each cluster, $(U'', V'', W'')$, relative to the family velocity. The position of each cluster relative to the family center in a coordinate system moving with the Local Standard of Rest is $(x'', y'', z'') = (x, y, z) - (x, y, z)_{\text{family}}$. This allows us to calculate the component of a cluster's velocity pointing toward or away from the family center,

$$v_{\exp} = (U'', V'', W'') \cdot \frac{(x'', y'', z'')}{\|(x'', y'', z'')\|} \quad (1)$$

We call these expansion velocities, as the velocity is positive for most clusters in a family. The median expansion velocities for the clusters of each family are given in Table 1, where the upper and lower bounds represent the spread in expansion velocities for a given family and are not considered uncertainties.

We explored the sensitivity of $D_{\text{compact}}$ and $t_{\text{compact}}$ to the choice of mass model and adopted solar peculiar motion. For the choices listed above, we found that the systematic changes in these parameters were smaller than the statistical errors associated with uncertainties in the positions and velocities of the individual clusters.

One characteristic of the family's trajectories that did change significantly with adopted solar peculiar motion (but not mass model) was the radial drift of the family members. We calculated the radial Galactocentric motion,

$\Delta R$ (pc), for each cluster (or its progenitor cloud) over the period $t = -60$ to $0$ Myr. For our adopted values of ($U_\odot$, $V_\odot$, $W_\odot$) = (11.1, 12.24, 7.25) km s$^{-1}$, the M6 and $\alpha$Per families moved radial outward by similar, positive values of $\Delta R = +420.3\,^{+13.4}_{-10.2}, +374.1\,^{+21.8}_{-60.3}$ pc. In contrast, the Cr135 family moved radially inward by $\Delta R = -247.6\,^{+41.1}_{-51.2}$ pc. These values shifted by about +500 pc when we used ($U_\odot$, $V_\odot$, $W_\odot$) = (10.0, 5.2, 7.2) km s$^{-1}$ and by $-200$ pc when assuming ($U_\odot$, $V_\odot$, $W_\odot$) = (10.1, 15.4, 7.8) km s$^{-1}$. This difference is presumably due to uncertainties in $V_\odot$.

**Inferring stellar masses, initial gas masses, and past supernovae counts**

To obtain a mass estimate for the three families of clusters, we construct color-absolute magnitude diagrams (CAMDs) using the stars' *Gaia* G-band magnitudes and parallaxes to calculate their absolute magnitudes and using G – G$_{RP}$ as their colors. For each cluster CAMD, we match an isochrone from a selection of PARSEC models[53] using the ages and extinction values reported for each cluster[5] and assuming a solar metallicity of $Z_\odot$ = 0.0158. For a given cluster, we can use the isochrone to estimate the mass of each star. To correct for the observational incompleteness of *Gaia* at both the low and high stellar mass regimes, we fit each mass distribution to an initial mass function (IMF)[28]. Since *Gaia* DR3 is expected to be complete in the G-band for magnitudes between 12 and 17[13], we convert these limits to absolute magnitudes based on a given cluster's distance and identify the expected completeness limits in stellar mass from the cluster's corresponding isochrone. We randomly draw 1,000 IMFs from a parent sample of 49,900 IMFs that range from 10 M$_\odot$ to 5,000 M$_\odot$ and are separated by steps of 0.1 M$_\odot$. For each cluster, we select the IMF that best matches the observed total mass in the computed complete mass range[54]. The total mass of the selected IMF is then computed as the total mass of the cluster. We repeat this process 100 times for each cluster to calculate the 16th, 50th, and 84th percentiles in total cluster mass. We compare 35 clusters in our sample that crossmatch to another work[55] which estimated cluster masses. We find that the masses correlate well between our work and this previous work, however our masses are systematically shifted higher by a factor of 30%. This shift, however, is well within the corresponding statistical uncertainties[55]. We compute the summed stellar mass of a cluster family by summing 1,000 randomly generated samples from the individual distributions of mass for each cluster of a given family, assuming that the mass uncertainty of a given cluster is distributed normally.

We estimate the initial gas mass (M$_{gas}$) of each family as

$$M_{gas} = \frac{M_{stars}}{SFE} - M_{stars} \quad (2)$$

where $M_{stars}$ is the computed stellar mass of a family and SFE is the star formation efficiency ranging between 0.5 and 1.5%. The number of massive stars (> 8 M$_\odot$) that formed in a cluster and have since exploded as supernovae is calculated by taking the best-matching IMF model and summing the number of stars greater than the most massive star currently present as indicated by a cluster's isochrone (assuming that this value is greater than 8 M☉). These values are summed for each family and reported in Table 1. The uncertainties in these values are determined using the same methodology applied to the stellar masses. The masses, SNe counts, and their corresponding uncertainties are reported for each cluster in the source data table included in the Supplementary Information. We do not calculate the masses or SNe output of the 18 YLAs in our sample since their memberships are significantly incomplete[15]. This omission does not impact our total family mass calculations, as the YLAs constitute only 0.01% of the total stars in our sample. We note that our statistical uncertainties do not fully capture all the uncertainties in our estimates of cluster stellar mass and SNe counts. Despite our confidence in the integrity of our adopted cluster catalog, which is likely to be mostly free from false-positive stellar members, it is probable that due to the limitations of the HDBSCAN clustering algorithm, an unspecified portion of genuine stellar members might be missing from a given cluster[14]. Therefore, it is possible that our computed stellar masses and SNe counts are lower limits.

We select the clusters of the Alpha Persei family that could have contributed to the formation of the Local Bubble (LB) via their SNe output. We do this by first visually selecting the clusters composing the regions of UCL and LCC. The clusters of UCL include V1062-Sco (UPK 640), UPK 606, OC 0666, HSC 2816, OCSN 92, HSC 2733, HSC 2636. The clusters of LCC include HSC 2523, HSC 2505, and HSC 2468. Although IC 2602 and Platais 8 are not formal LCC members, their present-day proximity to LCC and co-movement warrant their inclusion. Additionally, we incorporate three older clusters (>10 Myr) from the USCO region (HSC 2907, Theia 67, and HSC 2919), due to their proximity to UCL and LCC approximately 10 – 15 Myr ago, potentially linking them to the LB's expansion. The combined supernovae output of these 16 clusters is $11^{+4}_{-4}$ SNe.

**Calculating the age and energy input of supernovae-driven dust shells**

We use an analytic model of an expanding, thin, adiabatic shell in a uniform medium driven by SNe from an interior stellar association to estimate both the dynamical age of GSH 238 and the necessary energy input via SNe necessary to drive its expansion[32]. Expressing these equations in terms of directly observable quantities[56] yields:

$$t_7 = \left(\frac{R}{97 \text{ pc}}\right) \times \left(\frac{v}{5.7 \text{ km s}^{-1}}\right)^{-1} \quad (3)$$

$$N_* = n_e \times \left(\frac{R}{97 \text{ pc}}\right)^2 \times \left(\frac{v}{5.7 \text{ km s}^{-1}}\right)^3 \quad (4)$$

The quantity $t_7$ is the dynamical age of the shell in units of 10 Myr and $N_*$ is the number of massive stars (> 8 M☉) formed in the stellar association with the assumption that each input $10^{51}$ ergs of mechanical energy via a supernovae explosion. The quantity $n_e$ is the particle electron density of the ambient ISM. We assume a range of

1 – 3 particles cm$^{-3}$. The expansion velocity of GSH 238 from HI data has been measured as $v$ = 8 km s$^{-1}$ [19]. Using the 3D dustmap displayed in Figure 2, we obtain a semi-major axis of 500 pc measured from the edge of the Local Bubble model (in the direction of GSH 238) to the far end of the shell. The semi-minor axis is 300 pc; therefore, we consider a range of radii values of $R$ = 300 – 500 pc. We compute $N_*$ and $t_7$ 10000 times, each time using values drawn randomly from the ranges for $n_e$ and $R$, deriving a super-shell age of 29.4 ± 4.3 Myr with number of supernovae given by $N_*$ = 96 ± 40.

We also assess the consistency of the dynamical age $t_7$ and the number of supernovae $N_*$ for the Local Bubble (LB) with established expectations. Adopting previously determined parameters[18] — expansion velocity ($v$ = 6.3 – 7.2 km s$^{-1}$), radius (R = 159 – 171 pc), and ambient particle density ($n_e$ = 1.7 – 4.2) particles cm$^{-3}$ [18] — we calculate the LB's dynamical age to be $t_7$ = 14.4 ± 0.7 Myr and estimate the number of contributing supernovae to be $N_*$ = 14.2 ± 4.0, which is consistent with the expansion age found from a previous work and the number of supernovae expected to have exploded from the UCL/LCC and nearby clusters (11± 4 SNe).

Although the M6 family's dust cavity is not identified as a singular, expanding shell, by inverting equations 3 and 4, we can derive the current radius and expansion speed of a shell generated by the M6 family. Adopting the average age of the M6 family ($32.4^{+13.8}_{-10.5}$ Myr) as $t_7$, the calculated number of supernovae ($124^{+11}_{-10}$) as $N_*$, and $n_e$ = 1 – 3 particles cm$^{-3}$, we find that a shell produced by the M6 family would have a radius, $R$ = 473 ± 41 pc, and expansion velocity, $v$ = 7.9 ± 0.6 km s$^{-1}$.

It should be noted that the expansion velocity and radius of the shells are derived using a mean power delivered by SNe of $P_{SN} \sim$ (6.3 x 10$^{35}$ ergs s$^{-1}$) × ($N_* E_{51}$) [32]. This power assumes a constant rate of $N_*$ explosions spread out over a typical cluster lifetime of $t \sim$ 5 x 10$^7$ years. In the case of GSH 238 and the Cr135 family, our estimate of the massive star population indicates there are 108 stars greater than 8 solar masses, of which 63 have already exploded and 45 have yet to explode. Using the above formalism and the needed number of SNe to produce GSH 238, we calculate $P_{SN}$ = (6.3 x 10$^{35}$ ergs s$^{-1}$) x (96) = 6.0 x 10$^{37}$ ergs s$^{-1}$. For comparison, we divide the output mechanical energy from the SNe of the Cr135 family (63 x 10$^{51}$ ergs s$^{-1}$) by an assumed mean lifetime of 20 Myr (6.3 x 10$^{14}$ seconds) for the family, which yields $P_{SN}$= 1.0 x 10$^{38}$ ergs s$^{-1}$, in rough agreement with but a factor of 1.7 higher than the previous computed value. Lowering the fraction of SN energy that goes into thermalization of the superbubble interior, increasing the timescale to account for the extended duration of cluster formation, or modifying the assumption of a constant supernova luminosity might reduce our calculated SN power to the numbers assumed in the above formulation. This will be the subject of future investigations.

## Data Availability

The table of 272 clusters constructed and analyzed in this work, along with the interactive versions of Figure 1 and 2, are publicly available at the Harvard Dataverse: The table of clusters can be downloaded at the following link and includes the clusters' family memberships, mass estimates, and estimated SNe counts recovered in this work: https://doi.org/10.7910/DVN/VYBZQS. The interactive version of Figure 1 can be downloaded at the following link: https://doi.org/10.7910/DVN/IDJGDW. The interactive version of Figure 2 can be downloaded at the following link: https://doi.org/10.7910/DVN/MYWA1K.

## Code Availability

The code used for the analysis is available from CS upon reasonable request. Publicly available software libraries were used, including *galpy*[23], *astropy*[57], and *hdbscan*[24]. The static figures were produced using the *matplotlib*[58] and *healpy*[59] libraries, and the 3D interactive figures were made using the *plotly python* library. The interactive all-sky figure was developed using the *python World Wide Telescope (pywwt)* library.

## Methods References

# Acknowledgements

We thank Bruce Elmegreen for his helpful discussion of the results presented in this work. We also thank Jon Carifio for his assistance in developing the all-sky interactive figure. JA was co-funded by the European Union (ERC, ISM-FLOW, 101055318). Views and opinions expressed are, however, those of the author(s) only and do not necessarily reflect those of the European Union or the European Research Council. Neither the European Union nor the granting authority can be held responsible for them. SR acknowledges funding by the Austrian Research Promotion Agency (FFG, https://www.ffg.at/) under project number FO999892674. JG acknowledges funding by the Austrian Research Promotion Agency (FFG) under project number 873708. JG gratefully acknowledges the Collaborative Research Center 1601 (SFB 1601 sub-project A5) funded by the Deutsche Forschungsgemeinschaft (DFG, German Research Foundation) – 500700252. AG and CZ acknowledge support by NASA ADAP grant 80NSSC21K0634 "Knitting Together the Milky Way: An Integrated Model of the Galaxy's Stars, Gas, and Dust." CZ acknowledges that support for this work was provided by NASA through the NASA Hubble Fellowship grant #HST-HF2-51498.001 awarded by the Space Telescope Science Institute (STScI), which is operated by the Association of Universities for Research in Astronomy, Inc., for NASA, under contract NAS5-26555. This project was partly developed at the Lorentz Center workshop "Mapping the Milky Way", held in Leiden, Netherlands, in February 2023. The authors acknowledge Interstellar Institute's program "II6" and the Paris-Saclay University's Institut Pascal for hosting discussions that nourished the development of the ideas behind this work. This work has made use of data from the European Space Agency (ESA) mission Gaia https://www.cosmos.esa.int/ gaia, processed by the Gaia Data Processing and Analysis Consortium (DPAC, https://www.cosmos.esa.int/web/gaia/dpac/consortium). Funding for the DPAC has been provided by national institutions, in particular the institutions participating in the Gaia Multilateral Agreement.

# Author Contributions

CS led the work and wrote the majority of the text. All authors contributed to the text. CS, JA, RB, and SR led the data analysis, aided by JG and ELH. CS, JA, and RB led the interpretation of the results, aided by NMR, JG, SM, RK, and CZ. CS, NMR, ELH, and SR led the compilation of the stellar cluster catalog. CS, JA, and AG led the visualization efforts.

## Competing Interests

The authors declare that they have no competing financial interests.

## Author Information

Correspondence and requests for materials should be addressed to CS (email: cameren.swiggum@univie.ac.at).

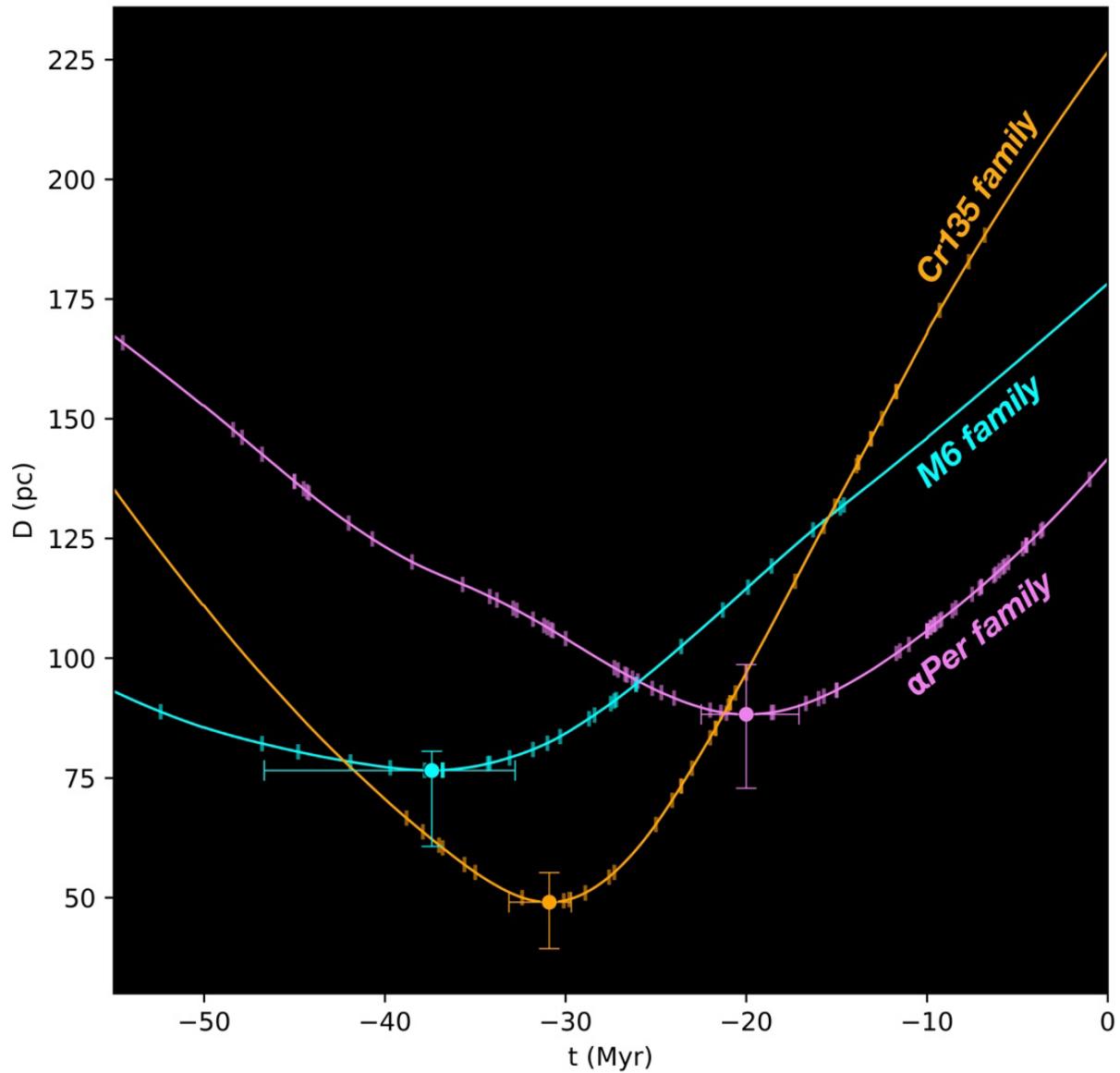

**Extended Data Figure 1: Sizes of each family computed for each timestep of their orbits.** Dots demarcate the timesteps and sizes of most compactness for a family. The error bars represent the 16th and 86th percentiles for the most compact size and the associated time, as determined from 1000 bootstrap samples of clusters randomly selected from a given family. The ticks along each family's size evolution profile indicate the birth times of individual clusters within that family.

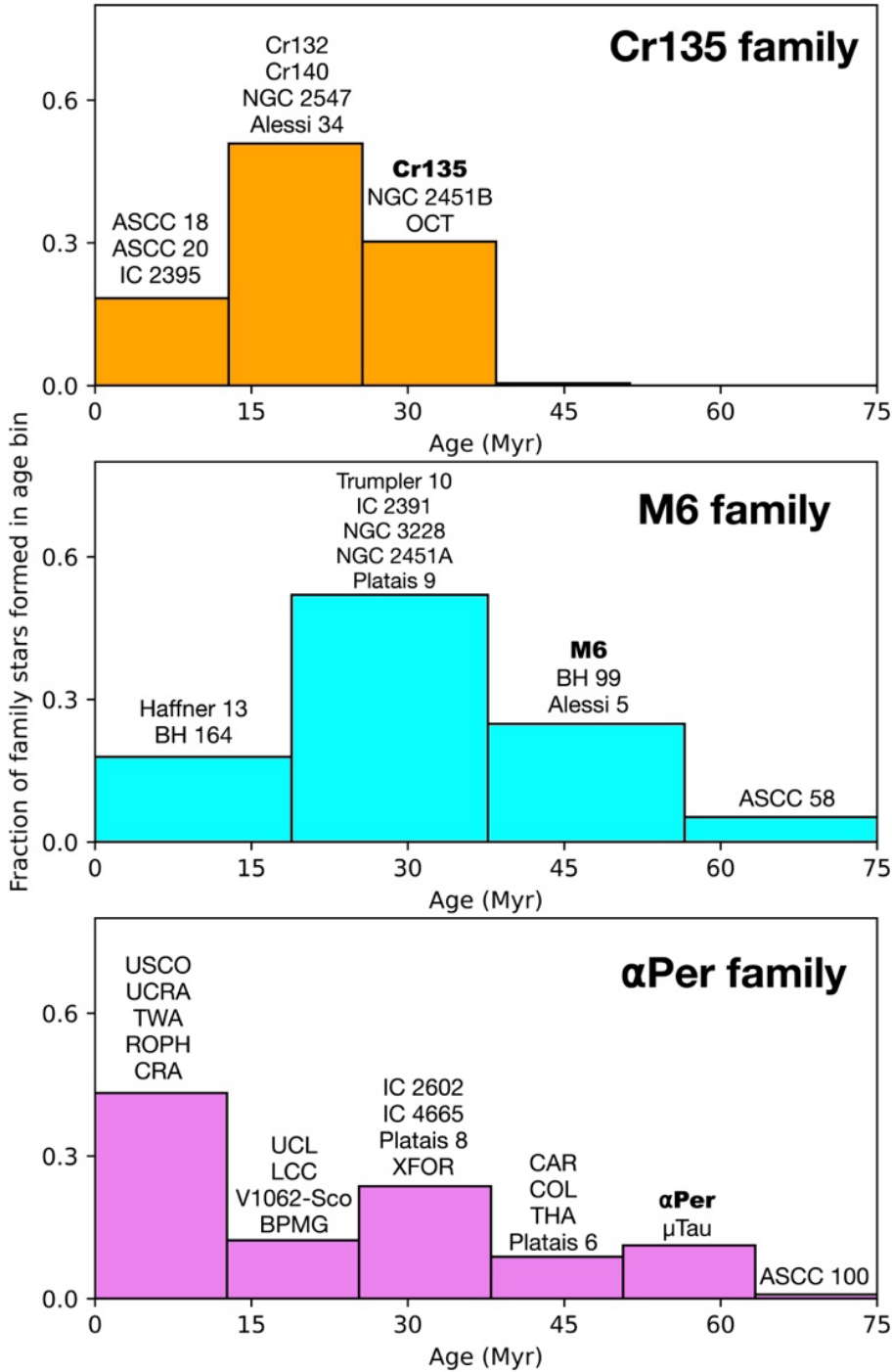

**Extended Data Figure 2: Histograms of Cluster Family Age Distributions.** Each histogram shows the number of clusters born as a function of lookback time for each cluster family. The bin widths were chosen based on the median age error. The y-axis shows the fraction of a family's stars in each age bin. The names of notable clusters are listed above their respective age bins.